\documentclass[aps, twocolumn,superscriptaddress,preprintnumbers,amssymb]{revtex4-1}
\usepackage{graphicx}
\usepackage{dcolumn}
\usepackage{bm}

\begin{document}

\preprint{ver6}

\title{Elastoresistance measurements on CaKFe$_4$As$_4$ and KCa$_2$Fe$_4$As$_4$F$_2$ with the Fe site of $C_{2v}$ symmetry}



\author{Taichi Terashima}
\email{TERASHIMA.Taichi@nims.go.jp}
\affiliation{International Center for Materials Nanoarchitectonics, National Institute for Materials Science, Tsukuba 305-0003, Japan}
\author{Yoshitaka Matsushita}
\affiliation{Research Network and Facility Services Division, National Institute for Materials Science, Tsukuba, Ibaraki 305-0047, Japan}
\author{Hiroyuki Yamase}
\affiliation{International Center for Materials Nanoarchitectonics, National Institute for Materials Science, Tsukuba 305-0003, Japan}
\author{Naoki Kikugawa}
\author{Hideki Abe}
\affiliation{Center for Green Research on Energy and Environmental Materials, National Institute for Materials Science, Tsukuba, Ibaraki 305-0003, Japan}
\author{Motoharu Imai}
\author{Shinya Uji}
\affiliation{Research Center for Functional Materials, National Institute for Materials Science, Tsukuba 305-0003, Japan}
\author{Shigeyuki Ishida}
\author{Hiroshi Eisaki}
\author{Akira Iyo}
\email{iyo-akira@aist.go.jp}
\author{Kunihiro Kihou}
\author{Chul-Ho Lee}
\affiliation{National Institute of Advanced Industrial Science and Technology (AIST), Tsukuba, Ibaraki 305-8568, Japan}
\author{Teng Wang}
\author{Gang Mu}
\email{mugang@mail.sim.ac.cn}
\affiliation{State Key Laboratory of Functional Materials for Informatics, Shanghai Institute of Microsystem and Information Technology, Chinese Academy of Sciences, Shanghai 200050, China}
\affiliation{CAS Center for Excellence in Superconducting Electronics (CENSE), Shanghai 200050, China}


\date{\today}

\begin{abstract}
We report resistance and elastoresistance measurements on (Ba$_{0.5}$K$_{0.5}$)Fe$_2$As$_2$, CaKFe$_4$As$_4$, and KCa$_2$Fe$_4$As$_4$F$_2$.
The Fe-site symmetry is $D_{2d}$ in the first compound but $C_{2v}$ in the latter two, which lifts the degeneracy of the Fe $d_{xz}$ and $d_{yz}$ orbitals.
The temperature dependence of the resistance and elastoresistance is similar between the three compounds.
Especially, the [110] elastoresistance is enhanced with decreasing temperature irrespective of the Fe-site symmetry.
This appears to be in conflict with recent Raman scattering studies on CaKFe$_4$As$_4$, which suggest the absence of nematic fluctuations.
We consider possible ways of reconciliation and suggest that the present result is important in elucidating the origin of in-plane resistivity anisotropy in iron-based superconductors.
\end{abstract}


\maketitle


\section{introduction}

Parent compounds of iron-based superconductors typically exhibit a tetragonal-to-orthorhombic structural phase transition at $T_s$ and a stripe-type antiferromagnetic transition at $T_N$ ($\leqslant T_s$) \cite{Kamihara08JACS, Cruz08nature, Rotter08PRB}. 
In spite of the tiny orthorhombicity $\delta = (a-b)/(a+b) = 2-4 \times 10^{-3}$ \cite{Cruz08nature, Rotter08PRB}, noticeable in-plane anisotropy appears in various electronic properties below $T_s$ as revealed by e.g. resistivity\cite{Chu10Science}, optical conductivity \cite{Dusza11EPL, Nakajima11PNAS}, inelastic neutron scattering (INS) \cite{Harriger11PRB, Ewings11PRB}, and NMR measurements \cite{Fu12PRL}.
We especially note that angle-resolved photoemission spectroscopy (ARPES) measurements show that the degenerate $d_{xz}$ and $d_{yz}$ levels of Fe in the tetragonal phase split considerably below $T_s$ \cite{Yi11PNAS}.
Those observations suggest that the transition at $T_s$ is electronically driven, and hence it is regarded as an electronic nematic transition (see e.g. \cite{Fernandes14NatPhys, BOHMER16CRP, Gallais16CRP} for a review). 
However, which electronic degrees of freedom, spin or orbital, are responsible for the nematic transition is still highly debated \cite{Yamakawa16PRX, Fernandes14NatPhys}.
The primary order parameter of the nematic phase would be deduced from the difference in spin fluctuations at $Q$ = ($\pi$, 0) and (0, $\pi$) (1-Fe unit cell) in the former case, while it would be the difference in the occupation between the $d_{xz}$ and $d_{yz}$ orbitals in the latter \cite{Fernandes14NatPhys}.

Nematic fluctuations in the tetragonal phase above $T_s$ are probed by various techniques.
The elastoresistance, which we report on in this paper, refers to the change in the resistance $\Delta R/R$ as the strain $\epsilon$ is applied and is defined as $m = \mathrm{d}(\Delta R/R)/\mathrm{d}\epsilon$, which is a measure of resistivity anisotropy between the strain direction and the perpendicular direction induced by the strain.
The elastoresistance with strain applied along the tetragonal [110] direction, which becomes the $a$ or $b$ axis in the orthorhombic phase below $T_s$, is assumed to be a proxy for the nematic susceptibility \cite{Chu12Science}.
Raman scattering in $B_{1g}$ symmetry (1-Fe unit cell) can detect nematic fluctuations and hence can be used to derive the nematic susceptibility \cite{Yamase11PRB, Gallais13PRL, Yamase13PRB_Raman, Kontani14PRL, Yamase15NJP, Khodas15PRB, Karahasanovic15PRB, Kretzschmar16NatPhys, Thorsmolle16PRB}.
The shear modulus can also be related to the nematic susceptibility \cite{Fernandes10PRL}.
Nematic susceptibilities in 1111-, 122-, and 11-iron-based superconductors (or parent compounds) estimated by these different techniques are broadly consistent (when a comparison can be made), exhibiting strong enhancement as $T \rightarrow T_s$ (or $T \rightarrow 0$ in the case of moderately overdoped compounds in which the structural transition does not occur down to absolute zero) \cite{Chu12Science, Kuo13PRB, Kuo14PRL, Kuo16Science, Liu16PRL, Tanatar16PRL, Hosoi16PNAS, Gu17PRL, Chauviere10PRB, Gallais13PRL, Gnezdilov13PRB, Thorsmolle16PRB, Kretzschmar16NatPhys, Massat16PNAS, Kaneko17PRB, Fernandes10PRL, Goto11JPSJ, Yoshizawa12JPSJ, Bohmer14PRL, Bohmer15PRL}.

\begin{figure}
\includegraphics[width=8.6cm]{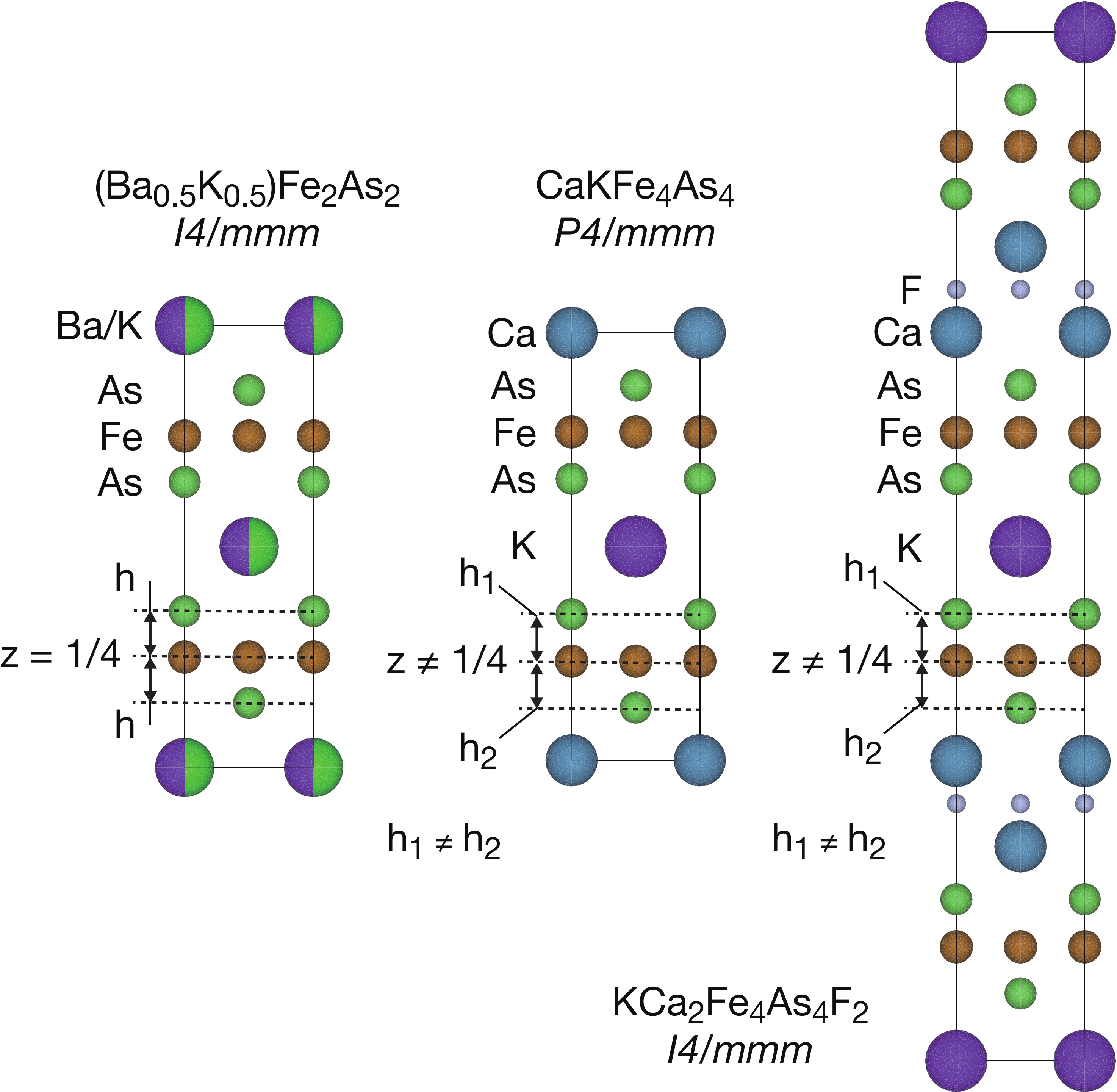}%
\caption{\label{struct}Crystal structures of (Ba$_{0.5}$K$_{0.5}$)Fe$_2$As$_2$, CaKFe$_4$As$_4$, and KCa$_2$Fe$_4$As$_4$F$_2$ (from left to right, prepared using the software VESTA \cite{Momma11JAC}).
In (Ba$_{0.5}$K$_{0.5}$)Fe$_2$As$_2$, the Fe layers are located at $z$ = 1/4 (and equivalent), and the heights (distances) $h$ measured from the Fe layers of the As layers above and below are the same.
The symmetry of the Fe site is $D_{2d}$, where the $d_{xz}$ and $d_{yz}$ orbitals are degenerate.
In CaKFe$_4$As$_4$ and KCa$_2$Fe$_4$As$_4$F$_2$, the Fe layers are displaced from $z$ = 1/4, and the height $h_1$ of the As layers above and $h_2$ of the As layers below are different.
The symmetry of the Fe site is lowered to $C_{2v}$, where the $d_{xz}$ and $d_{yz}$ orbitals are no longer degenerate.
}
\end{figure}

In this context, it is intriguing that recent Raman scattering measurements on CaKFe$_4$As$_4$ \cite{Iyo16JACS} fail to observe nematic fluctuations \cite{Jost18PRB, Zhang18PRB}, despite the fact that the [110] elastroresistance is enhanced with decreasing temperature \cite{Meier16PRB}.
The authors of \cite{Zhang18PRB} ascribe the absence of nematicity in CaKFe$_4$As$_4$ to the reduced symmetry of the Fe site (see Fig.~\ref{struct}).
The Fe site symmetry in 1111-, 122-, and 11-type iron-based superconductors (or parent compounds) is $D_{2d}$, where the $d_{xz}$ and $d_{yz}$ orbitals are degnerate.
On the other hand, the Fe site symmetry in CaKFe$_4$As$_4$ is $C_{2v}$ as explained in Fig.~\ref{struct}, where the $d_{xz}$ and $d_{yz}$ orbitals are no longer degenerate.
The authors argue that this would result in a static antiferroquadrupolar (AFQ) order and preclude ``Pomeranchuk-like fluctuations'' \cite{Zhang18PRB}.

The present study was motivated by this contradiction.
We perform elastoresistance measurements on (Ba$_{0.5}$K$_{0.5}$)Fe$_2$As$_2$, CaKFe$_4$As$_4$, and KCa$_2$Fe$_4$As$_4$F$_2$ \cite{Wang16JACS}.
As Fig.~\ref{struct} shows, CaKFe$_4$As$_4$ is an ordered hybrid of two 122 compounds CaFe$_2$As$_2$ and KFe$_2$As$_2$, while KCa$_2$Fe$_4$As$_4$F$_2$ is that of a 1111 compound CaFeAsF \cite{Matsuishi08JACS} and KFe$_2$As$_2$.
The three compounds have the same doping level of 0.25 hole per Fe, corresponding to a slightly overdoped region, and remain tetragonal down to absolute zero temperature.
The Fe-site symmetry of the first one is $D_{2d}$, while that of the latter two is $C_{2v}$ (Fig.~\ref{struct}).
The first one is measured for comparison.
We confirm the enhancement of the [110] elastoresistance in CaKFe$_4$As$_4$ reported in \cite{Meier16PRB}, and further find that the [110] elastoresistance is also enhanced in KCa$_2$Fe$_4$As$_4$F$_2$.
Implications of the results will be discussed.


%

\section{experiments}

Single crystals of (Ba$_{0.5}$K$_{0.5}$)Fe$_2$As$_2$ and CaKFe$_4$As$_4$ were prepared in Tsukuba using KAs and FeAs as flux, respectively, and were thoroughly characterized as described in \cite{Kihou16JPSJ, Ishida19npjQM}.
For the former, crystals were picked up from the same growth batch as the $x$ = 0.51 sample of \cite{Kihou16JPSJ}, but considering the known composition variation of a few percent from crystal to crystal within the same growth batch, we round the composition off to 0.5.
Single crystals of KCa$_2$Fe$_4$As$_4$F$_2$ were grown in Shanghai using KAs as flux and were thoroughly characterized as described in \cite{Wang19JPCC}.

For electrical resistance and elastoresistance measurements, [110]- and [100]-oriented samples were cut from grown crystals.
To gain the full strain transmission in elastoresistance measurements \cite{Chu12Science, Kuo16Science}, samples with a thickness less than 50 $\mu m$ and a length larger than 1 mm were used. 
X-ray diffraction was employed to determine the crystal axes.
Electrical contacts were spot-welded or made using silver paste.
Elastoresistance was measured using a piezostack in a way similar to \cite{Chu12Science}.
A sample and a strain gauge were glued on the surface of a piezostack, and the sample resistance $R$ was recorded as a function of the strain $\epsilon$ as the operating voltage of the piezostack was ramped up and down.
Instead of holding a constant temperature for each measurement, we collected $R$ vs $\epsilon$ data continuously, slowly cooling or warming the sample (typically 0.3 K/min), and corrected the data for the resistance variation due to the temperature variation to determine the elastoresistance (see Appendix for details).
Results reported below are for the longitudinal configuration; that is, the current $I$ was applied parallel to the strain $\epsilon$.
The elastoresistance coefficients $m_{[110]}$ and $m_{[100]}$ were measured with $I \parallel$ [110] and [100], respectively.
Nematic fluctuations under consideration enhance $m_{[110]}$ but not $m_{[100]}$.
Some previous studies determined values of 2$m_{66}$ and ($m_{11}$ - $m_{12}$) by using both longitudinal and transverse configurations \cite{Kuo13PRB} or more sophisticatedly by using a modified Montgomery method \cite{Kuo16Science} or transverse-resistivity configuration \cite{Shapiro16RSI}.
$m_{ij}$'s are components of the elastoresistance tensor \cite{Kuo13PRB}, and 2$m_{66}$ and ($m_{11}$ - $m_{12}$) represent the elastoresistance responses in the $B_{2g}$ and $B_{1g}$ symmetry channels (2-Fe unit cell), respectively \cite{Shapiro15PRB}.
$m_{[110]}$ and $m_{[100]}$ are practically proportional to 2$m_{66}$ and ($m_{11}$ - $m_{12}$) \cite{Chu12Science, Hosoi16PNAS, Ishida20PNAS}.

\section{results and discussion}

\begin{figure*}
\includegraphics[width=17.8cm]{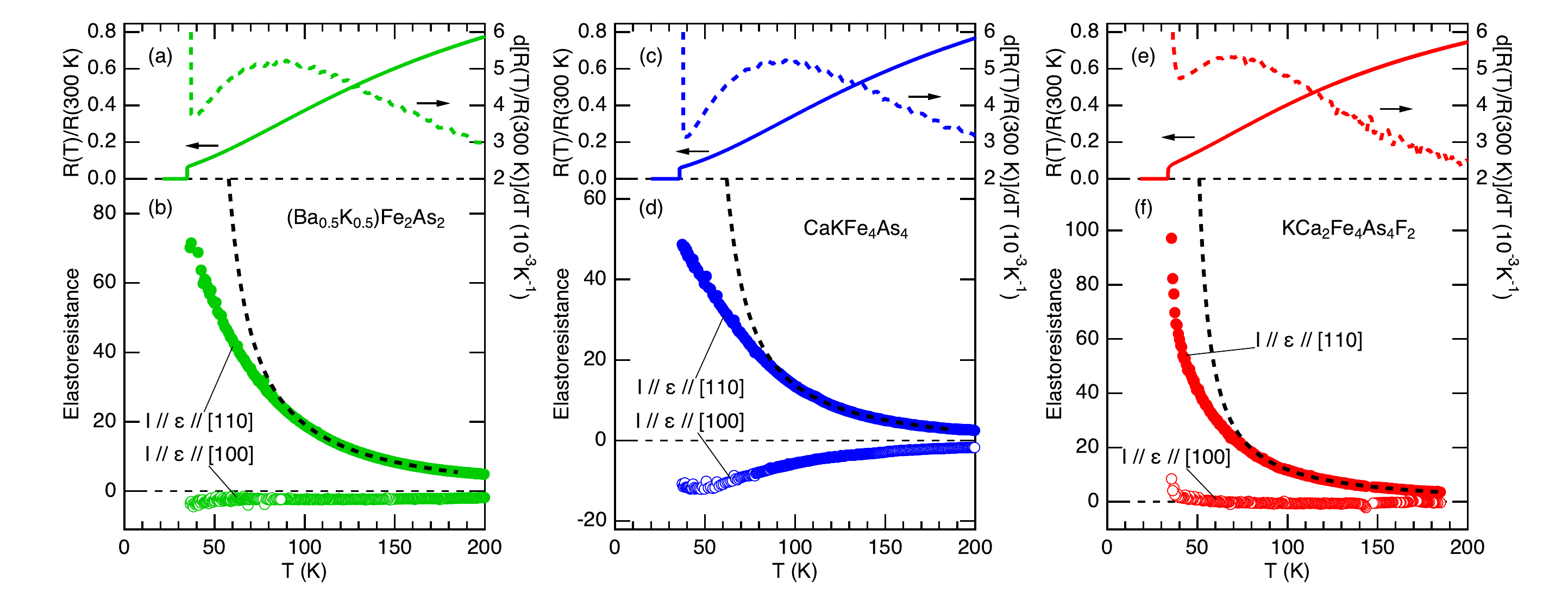}
\caption{\label{elastoR} Temperature dependence of resistance (a, c, e) and elastoresistance (b, d, f) for  (Ba$_{0.5}$K$_{0.5}$)Fe$_2$As$_2$, CaKFe$_4$As$_4$, and KCa$_2$Fe$_4$As$_4$F$_2$ (from left to right).  The dashed lines in (a, c, e) are the temperature derivatives.  In the elastoresistance measurements (b, d, f), the electrical current $I$ and the strain $\epsilon$ are parallel and applied parallel to [110] and [100].
The dashed lines show Curie-Weiss fits $m_{[110]} = C/(T-\theta) + m_0$ to the [110] data in the region between 100 and 185 K (see text).
}
\end{figure*}

\begin{table*}
\caption{\label{tab1} The second column shows resistivities at $T$ = 300 K of (Ba$_{0.5}$K$_{0.5}$)Fe$_2$As$_2$ and CaKFe$_4$As$_4$ estimated from measurements on four and five samples, respectively.  Resistivity was not determined for KCa$_2$Fe$_4$As$_4$F$_2$.  The rest of the columns show values determined for the [110] samples of Fig.~\ref{elastoR}: the residual resistivity ratio defined as $R$(300 K)/$R$(36 K), the superconducting transition temperature $T_c$ determined from the midpoint of the transition, the coefficients $C$, $\theta$, and $m_0$ determined from the Curie-Weiss fit to the [110] elastoresistance data (see text).}
\begin{ruledtabular}
\begin{tabular}{ccccccc}
 & $\rho$(300 K) ($\mu \Omega$ cm) & $R$(300 K)/$R$(36 K) & $T_c$ (K) & $C$ (K) & $\theta$ (K) & $m_0$ \\
(Ba$_{0.5}$K$_{0.5}$)Fe$_2$As$_2$ & 430(180) & 14 & 34.8 & 1280(20) & 44.3(5) & -3.6(1)\\
CaKFe$_4$As$_4$ & 350(190) & 17 & 35.9 & 890(20) & 49(1) & -3.7(2) \\
KCa$_2$Fe$_4$As$_4$F$_2$ & -- & 12 & 33.7 & 750(20) & 45(1) & -1.8(1) \\
\end{tabular}
\end{ruledtabular}
\end{table*}

Figure~\ref{elastoR} shows the temperature dependence of the resistance and elastoresistance for (Ba$_{0.5}$K$_{0.5}$)Fe$_2$As$_2$, CaKFe$_4$As$_4$, and KCa$_2$Fe$_4$As$_4$F$_2$. 
For each compound, the same sample was used to measure the temperature dependence of the resistance and [110] elastoresistance, while a different sample was used for the [100] elastoresistance (it is impossible to remove epoxy without damaging samples).
In addition, at least one more sample for each compound was measured to confirm the reproducibility of the [110] elastoresistance.
For the [110] samples used in the figure, the residual resistivity ratio defined as $R$(300 K)/$R$(36 K) and superconducting transition temperature $T_c$ determined from the midpoint of transition are summarized in Table~\ref{tab1}.
The large residual resistivity ratios and transition temperatures indicate the high quality of our samples.
Although the actual resistivity $\rho$ is not easy to accurately determine because of uncertainty in the sample dimensions and inhomogeneous current distribution, especially along the interlayer direction, resistivities at $T$ = 300 K of (Ba$_{0.5}$K$_{0.5}$)Fe$_2$As$_2$ and CaKFe$_4$As$_4$ were estimated from measurements on four and five samples, respectively, and are shown in the table as well.
The resistivity was not determined for KCa$_2$Fe$_4$As$_4$F$_2$ because of technical reasons.
The dashed lines in Figs.~\ref{elastoR}(b), (d), and (f) are Curie-Weiss fits $m_{[110]} = C/(T-\theta) + m_0$ to the [110] data in the temperature range between 100 and 185 K (the upper bound 185 K was set because it was the highest temperature for the elastoresistance measurements on KCa$_2$Fe$_4$As$_4$F$_2$).
The estimated parameters $C$, $\theta$, and $m_0$ are also listed in Table~\ref{tab1}.

The three compounds show a similar temperature dependence of resistance [Figs.~\ref{elastoR}(a), (c), and (e)]: the $R(T)$ curves are slightly concave at high temperatures and become convex at low temperatures.
The inflection point is 86, 93, and 69 K for (Ba$_{0.5}$K$_{0.5}$)Fe$_2$As$_2$, CaKFe$_4$As$_4$, and KCa$_2$Fe$_4$As$_4$F$_2$, respectively (see the derivative curves).
Furthermore, the derivative curves (dashed lines) of the former two compounds emphasize a particularly close similarity between them.

The measured [110] elastoresistances are enhanced considerably with decreasing temperature \textit{irrespective of the Fe-site symmetry}, while the [100] ones are much smaller and much less temperature dependent [Figs.~\ref{elastoR}(b), (d), and (f)].
The [110] elastoresistances can be described well by the Curie-Weiss formula down to $\sim$100 K (dashed lines) but deviate downward at lower temperatures.
Although the coefficient $C$ in (Ba$_{0.5}$K$_{0.5}$)Fe$_2$As$_2$ is considerably larger than those in the other two (Table~\ref{tab1}), its significance is unclear.
A quantitative analysis of $C$ would require an elaborate theory dealing with changes in the electronic structure and electron scattering due to an applied strain.
The [110] elastoresistance data for (Ba$_{0.5}$K$_{0.5}$)Fe$_2$As$_2$ is qualitatively consistent with the 2$m_{66}$ data for (Ba$_{0.6}$K$_{0.4}$)Fe$_2$As$_2$ \cite{Kuo16Science}.
The present fit parameter $\theta$ =  44.3(5) K (Table \ref{tab1}) is close to $\theta$ =  46.1(2.4) K reported in \cite{Kuo16Science}.
The [110] elastoresistance data for CaKFe$_4$As$_4$ is also qualitatively consistent with the 2$m_{66}$ data reported in \cite{Meier16PRB}.

\begin{figure}
\includegraphics[width=8.6cm]{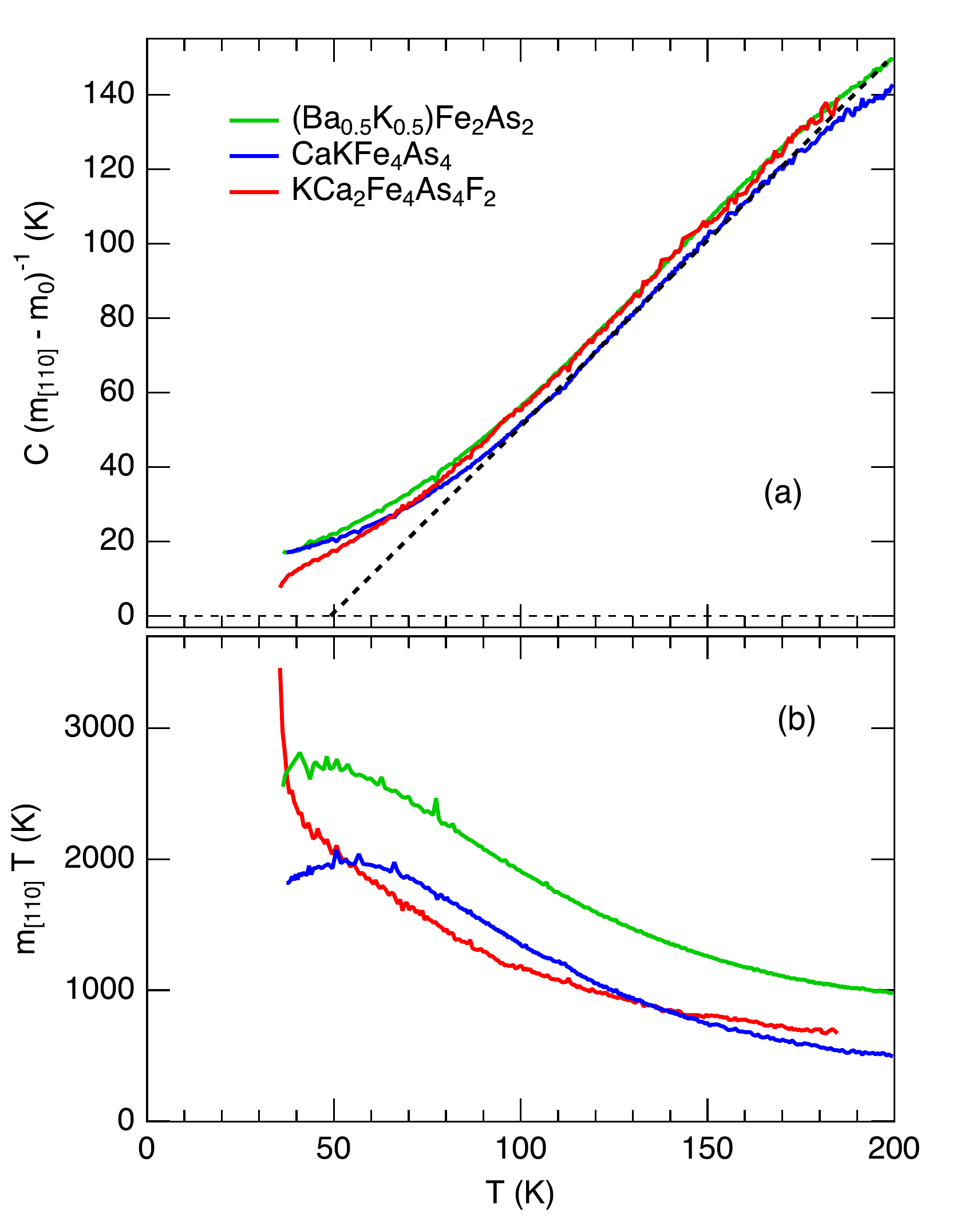}
\caption{\label{comp} (a) $C(m_{[110]} - m_0)^{-1}$ and (b) $m_{[110]} T$ as a function of temperature for (Ba$_{0.5}$K$_{0.5}$)Fe$_2$As$_2$, CaKFe$_4$As$_4$, and KCa$_2$Fe$_4$As$_4$F$_2$.
$m_{[110]}$ is the [110] elastoresistance shown in Figs.~\ref{elastoR}(b), (d), and (f), while $C$ and $m_0$ are the Curie-Weiss fit parameters listed in Table~\ref{tab1}.
The dashed line in (a) corresponds to the Curie-Weiss fit to the CaKFe$_4$As$_4$ data shown in Fig.~\ref{elastoR}(d).
}
\end{figure}

In order to compare the [110] elastoresistance data between the three compounds, we prepared two types of plots (Fig.~\ref{comp}).
Figure~\ref{comp}(a) shows $C(m_{[110]} - m_0)^{-1}$ as a function of temperature.
The [110] elastoresistance data of the three compounds show similar behavior: a Curie-Weiss behavior (i.e. straight line) in a limited temperature range, upward deviation at lower temperatures, and downward deviation at higher temperatures.
This is actually typical behavior observed in many other iron-based superconductors, especially nearly optimally doped compounds \cite{Kuo16Science, Hosoi16PNAS, Ishida20PNAS}.
Figure~\ref{comp}(b) shows $m_{[110]} T$ as a function of temperature.
This plot was motivated by the following idea:
for a local magnetic moment system, a plot of $\chi T$ vs $T$, where $\chi$ is the magnetic susceptibility, would give a flat line in the absence of interaction between the moments, but, generally, would show enhancement or suppression of $\chi T$ as temperature is lowered depending on whether the interaction is ferromagnetic or antiferromagnetic, and hence could give some insight into the interaction.
Figure~\ref{comp}(b) again emphasizes the particularly close similarity between (Ba$_{0.5}$K$_{0.5}$)Fe$_2$As$_2$ and CaKFe$_4$As$_4$: the two curves are almost parallel, indicating that the elastoresistance is similarly enhanced with decreasing temperature.
It is also interesting to note that for these two compounds $m_{[110]} T$ does not diverge but rather levels off or slightly decreases at low temperatures.
KCa$_2$Fe$_4$As$_4$F$_2$ behaves differently in this plot.
While the former two compounds are a 122 compound and a hybrid of 122's, the last one is a hybrid of a 122 and CaFeAsF.
CaFeAsF has a more two-dimensional electronic structure and stronger electronic correlations than 122 compounds \cite{Shein08JETPLett, Ma17SST, Terashima18PRX}, and KCa$_2$Fe$_4$As$_4$F$_2$ itself is also highly two-dimensional \cite{Wang16EPL, Wang19JPCC}.
These might be related to the difference.

It is unclear why the [110] elastoresistances are suppressed and deviate from the Curie-Weiss behavior at low temperatures.
The nematic susceptibility of (Ba$_{1-x}$K$_{x}$)Fe$_2$As$_2$ was also derived from shear modulus measurements in \cite{Bohmer14PRL}.
The obtained nematic susceptibility for K concentrations $x$ of 0.3 and higher cannot be described by a Curie-Weiss law with a constant Weiss temperature $\theta$.
The authors of \cite{Bohmer14PRL} introduce a temperature-dependent $\theta$ and estimate that it is negative for $x \geqslant 0.3$, contrary to the present result for $x$ = 0.5, and gets more negative as temperature is lowered.
The Curie-Weiss behavior of the nematic susceptibility was originally suggested from a phenomenological Ginzburg-Landau free-energy model which considers the nematic order parameter, elastic strain, and their linear coupling \cite{Chu12Science}.
The same model predicts that the structural transition should occur at a temperature higher than the Weiss temperature $\theta$, which is clearly inconsistent with the present experimental observations.
Inclusion of some other ingredients in the model or going beyond the mean-field level may be necessary to explain the observed suppression of the elastoresistances and the discrepancy between the elastoresistance and shear modulus data.
 
Our experimental results show that (Ba$_{0.5}$K$_{0.5}$)Fe$_2$As$_2$, CaKFe$_4$As$_4$, and KCa$_2$Fe$_4$As$_4$F$_2$ exhibit a similar temperature dependence of resistance and elastoresistance.
Especially, the similarity between the former two compounds is striking.
This indicates that the electronic structure and scattering mechanisms, which dominate the electrical conduction, are similar between the three compounds.
If the enhancement of the [110] elastoresistance in (Ba$_{0.5}$K$_{0.5}$)Fe$_2$As$_2$ is ascribed to nematic fluctuations, as widely believed, it therefore appears natural to ascribe the enhancement in the latter two compounds to nematic fluctuations as well.
However, this assumption is in conflict with the Raman scattering studies on CaKFe$_4$As$_4$ \cite{Jost18PRB, Zhang18PRB}, which did not observe nematic fluctuations as already mentioned.
We consider two possibilities to resolve this conflict in the following.

First, one can argue that, although nematic fluctuations exist in CaKFe$_4$As$_4$, they were missed in the Raman studies.
The authors of \cite{Zhang18PRB} suggest that nematic fluctuations in the charge/orbital sector may be precluded by the AFQ-order-like arrangement of the $d_{xz}$ and $d_{yz}$ orbitals in CaKFe$_4$As$_4$.
However, this does not necessarily mean that spin-driven nematicity, for which the degeneracy of the $d_{xz}$ and $d_{yz}$ orbitals is not a prerequisite, is precluded.
In addition, if the energy splitting of the $d_{xz}$ and $d_{yz}$ orbitals at each site is small enough, orbital physics may still play a role.
We note that Raman scattering due to nematic fluctuations in hole-doped (Ba$_{1-x}$K$_{x}$)Fe$_2$As$_2$ \cite{Wu17PRB} is already weaker than that in electron-doped Ba(Fe$_{1-x}$Co$_x$)$_2$As$_2$ \cite{Gallais13PRL, Kretzschmar16NatPhys}.
It is necessary to theoretically examine if the lowered Fe-site symmetry or the possible absence of nematic fluctuations in the charge/orbital sector in CaKFe$_4$As$_4$ may further suppress Raman scattering intensity (to the extent that scattering due to nematic fluctuations is missed).
Raman scattering measurements on KCa$_2$Fe$_4$As$_4$F$_2$ are also desirable.


Second, one can argue that enhanced elastoresistance does not necessarily indicate enhanced nematic fluctuations.
Fluctuations of some order parameter other than nematic ones may also enhance the elastoresistance if it couples to the lattice, breaks tetragonal symmetry, and brings about a unidirectional in-plane anisotropy in the electronic structure and/or scattering.
In the present case, mere stripe-type spin fluctuations without nematic correlations already suffice.
The enhancement of spin fluctuations with decreasing temperature is observed in NMR and INS measurements on CaKFe$_4$As$_4$ \cite{Cui17PRB, Ding18PRL, Iida17JPSJ,Xie18PRL} and KCa$_2$Fe$_4$As$_4$F$_2$ \cite{Luo20CPB}.
As a strain is applied, spin fluctuations at ($\pi$, 0) and (0, $\pi$) become inequivalent, which can result in resistivity anisotropy.
INS measurements on CaKFe$_4$As$_4$ and KCa$_2$Fe$_4$As$_4$F$_2$ under uniaxial pressure are desirable to assess the validity of this scenario.

Finally, it is highly debated whether the in-plane resistivity anisotropy in iron-based superconductors is due to anisotropic scattering or an anisotropic Fermi surface \cite{Chuang10Science, Nakajima12PRL, Ishida13PRL, Allan13NatPhys, Kuo14PRL, Mirri15PRL, Mirri16PRB, Tanatar16PRL, Valenzuela10PRL, Liang12PRL, Sugimoto14PRB, Gastiasoro14PRL, BreitkreizPRB14B, Schutt16PRB, Onari17PRB}. 
If the AFQ-order-like arrangement of the $d_{xz}$ and $d_{yz}$ orbitals in $C_{2v}$ Fe-site compounds \cite{Zhang18PRB} suppresses Fermi surface deformation due to an applied strain, the present observation of enhanced elastoresistance in CaKFe$_4$As$_4$ and KCa$_2$Fe$_4$As$_4$F$_2$ may suggest the dominant role of anisotropic scattering.

\section{coclusion}

We have observed that (Ba$_{0.5}$K$_{0.5}$)Fe$_2$As$_2$, CaKFe$_4$As$_4$, and KCa$_2$Fe$_4$As$_4$F$_2$ exhibit similar temperature dependence of resistance and elastoresistance.
The [110] elastoresistance is enhanced irrespective of the Fe-site symmetry.
The temperature dependence of the [110] elastoresistance can be described by a Curie-Weiss law down to $\sim$100 K but deviates downward at lower temperatures.
The origin of the deviation remains unclear.
The enhancement observed in the $C_{2v}$ Fe-site compounds CaKFe$_4$As$_4$ and KCa$_2$Fe$_4$As$_4$F$_2$ appears to be in conflict with the previous Raman studies \cite{Jost18PRB, Zhang18PRB}.
We suggested two possible ways of reconciliation: 
The first one assumes that nematic fluctuations were missed in the Raman studies, while the other assigns the enhanced elastoresistance to fluctuations other than nematic ones.
We also suggested that the present result might shed new light on the issue of the origin of the resistivity anisotropy in iron-based superconductors.
Further studies on $C_{2v}$ Fe-site compounds, not only experimental ones but also theoretical ones, are clearly called for.

\begin{acknowledgments}
We thank Suguru Hosoi for showing us unpublished data.
This work was supported in Japan by JSPS KAKENHI Grant Numbers 17K05556, 19H05823, and 19K15034.
This work was supported in China by the Youth Innovation Promotion Association of the Chinese Academy of Sciences (No. 2015187).
\end{acknowledgments}

\appendix*

\section{Elastoresistance measurements with slowly varying sample temperature}

\begin{figure}
\includegraphics[width=8.6cm]{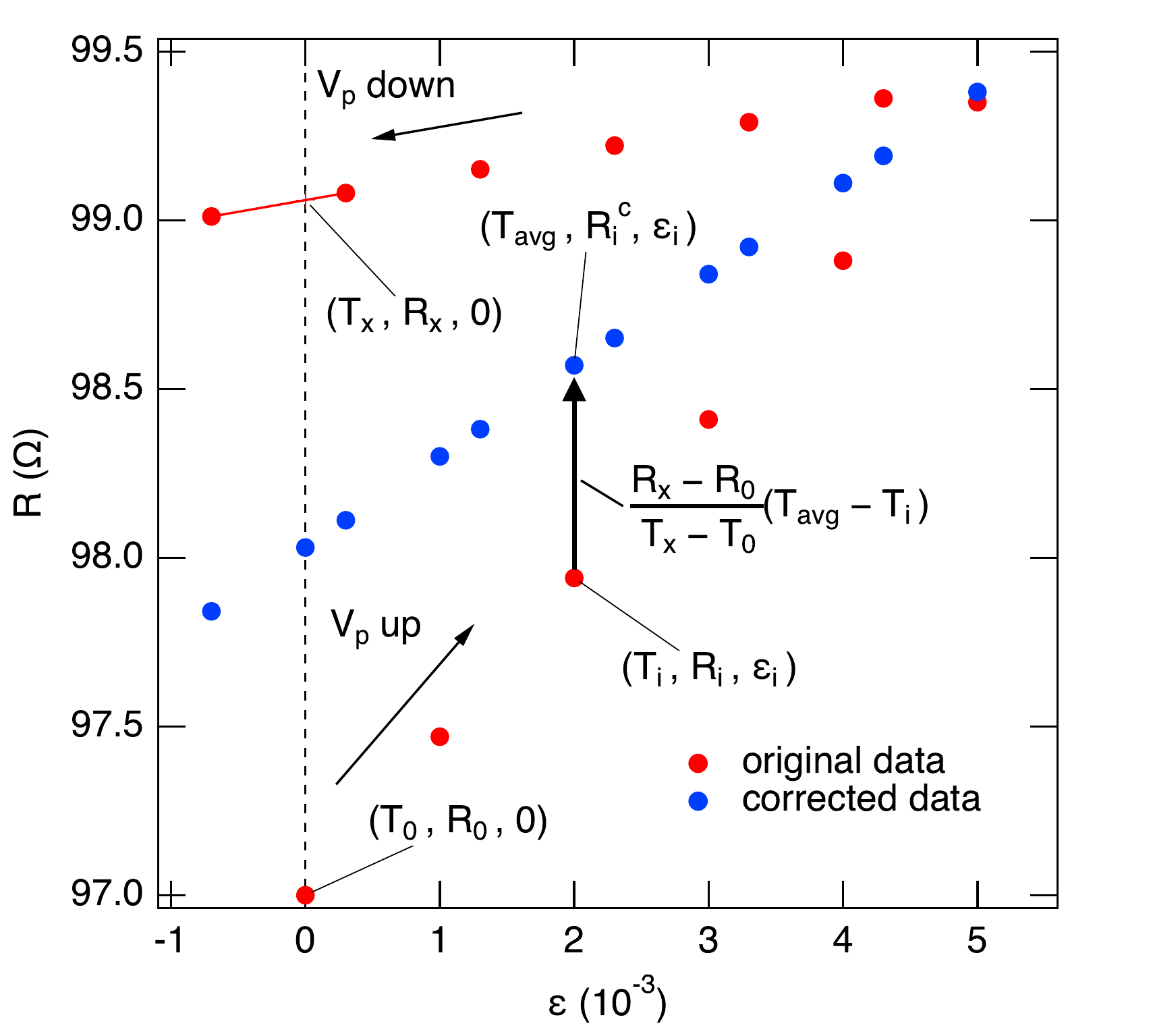}
\caption{\label{correction} Illustration of the temperature-variation correction.  See the text for explanation.
}
\end{figure}

\begin{figure}
\includegraphics[width=8.6cm]{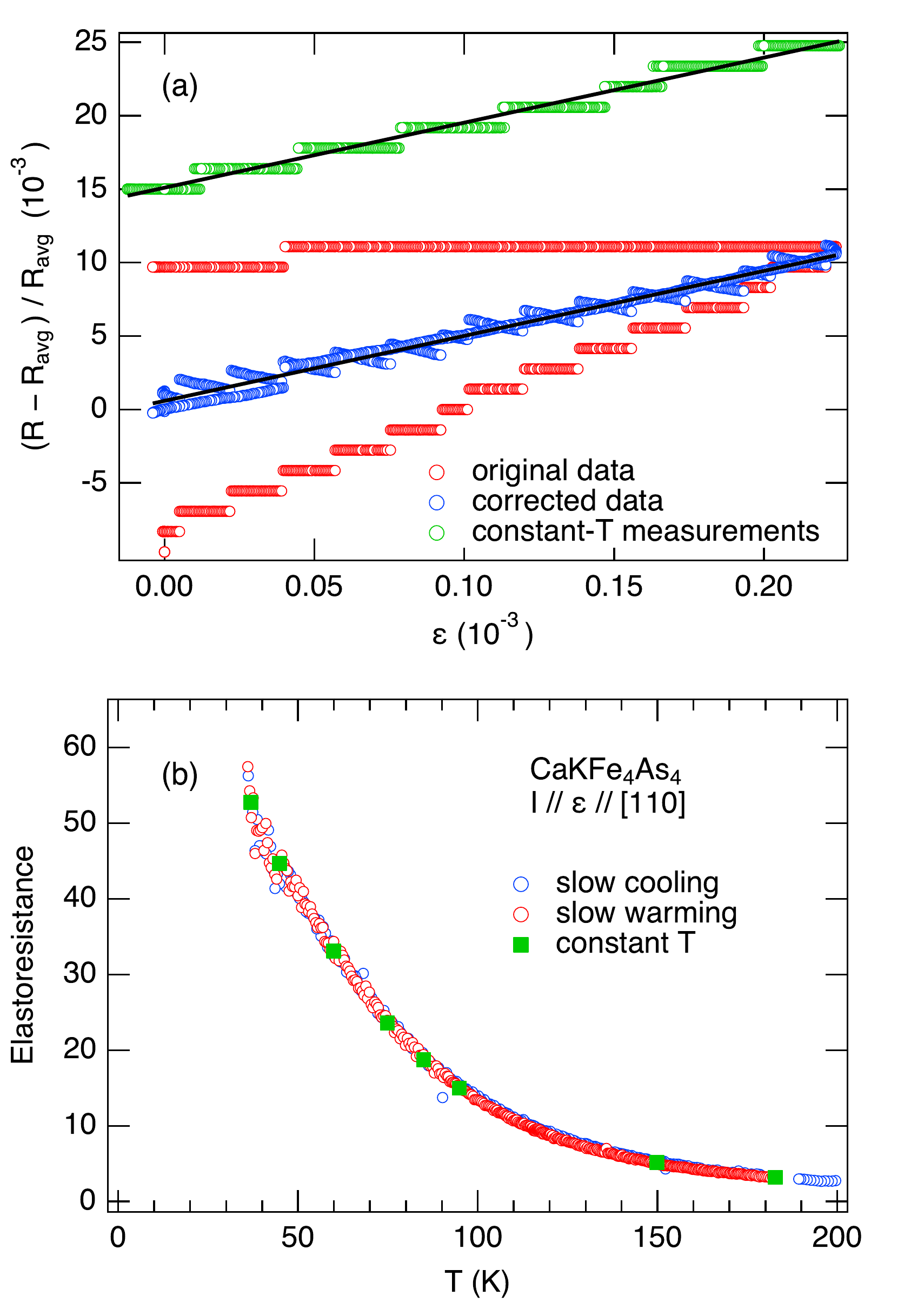}
\caption{\label{exampledata} (a) The red marks show original resistance vs strain data measured while the sample temperature was warmed slowly, while the blue marks show the corrected data for $T_{avg}$ = 45.036 K.
The green marks show data obtained with the sample temperature held constant at 44.870(2) K (vertically shifted for clarity).
The black lines are linear fits for two types of data, which give the same slope, i.e., the same elastoresistance value.
(b) The blue and red marks show elastoresistance values obtained using the present correction method while the sample was slowly cooled and warmed, respectively.
The green marks show elastoresistance values obtained while the sample temperature was stabilized at each temperature.
}
\end{figure}

We take a series of data ($T_{i}$, R$_{i}$, $\epsilon_{i}$) ($i$ = 0, \dots, $N$ - 1) as the operating voltage $V_{p}$ of a piezostack is ramped up and down while a sample is cooled or warmed slowly.
We measure the applied strain $\epsilon_{i}$ relative to the first data point for each $V_p$ cycle, so that $\epsilon_{0}$ = 0.
In order to obtain the elastoresistance, we correct the data for the temperature variation during the measurement as follows.
We interpolate the values of temperature $T_{x}$ and resistance $R_{x}$ at $\epsilon_{x}$ = 0 for the segment where $V_{p}$ is ramped down (see Fig. 4).
Then, we correct the resistance data $R_i$ as follows:
\begin{equation}
R_i^c = R_i + \frac{R_x - R_0}{T_x-T_0}(T_{avg}-T_i),
\end{equation}
where $T_{avg}=(T_0+T_x)/2$.
We plot $R_i^c$/$R_{avg}$ vs $\epsilon_i$, where $R_{avg}=(R_0+R_x)/2$, and fit a straight line to obtain a slope, which gives the elastoresistance.

Figure 5(a) shows an example.
The sample is a [110]-cut CaKFe$_4$As$_4$ sample different from the one used in Fig. 2.
The red circles show measured resistance vs strain data (the vertical axis is normalized using $R_{avg}$ as defined above).
The resistance was measured with a digital lock-in, and hence the resistance values are quantized due to the resolution of analog-to-digital conversion.
The measurement took 108 s, and the sample temperature varied from 44.795 K to 45.282 K.
The blue marks show the corrected resistance data at $T_{avg}$ = 45.036 K, and the black line is a linear fit, which gives a slope of 44.4(2).
For comparison, the green marks show resistance vs strain data measured holding a sample temperature of 44.870(2) K (vertically shifted for clarity).
The linear fit (black line) gives a slope of 44.4(2), the same as above.

Figure 5(b) shows elastoresistance vs temperature data.
The blue and red marks show elastoresistance values obtained using the above correction method while the sample was slowly cooled and warmed, respectively.
The green marks show elastoresistance values obtained while the sample temperature was stabilized at each temperature.
The excellent agreement between the three sets of data demonstrates the validity of our temperature-variation correction method.

\end{document}